\documentstyle[12pt,html,epsf,psfig]{article}

\begin{document}
\title{MATTERS OF GRAVITY, The newsletter of the APS Topical Group on 
Gravitation}
\begin{center}
{ \Large {\bf MATTERS OF GRAVITY}}\\ 
\bigskip
\hrule
\medskip
{The newsletter of the Topical Group on Gravitation of the American Physical 
Society}\\
\medskip
{\bf Number 21 \hfill Spring 2003}
\end{center}
\begin{flushleft}

\tableofcontents
\vfill
\section*{\noindent  Editor\hfill}

Jorge Pullin\\
\smallskip
Department of Physics and Astronomy\\
Louisiana State University\\
202 Nicholson Hall\\
Baton Rouge, LA 70803-4001\\
Phone/Fax: (225)578-0464\\
Internet: 
\htmladdnormallink{\protect {\tt{pullin@phys.lsu.edu}}}
{mailto:pullin@phys.lsu.edu}\\
WWW: \htmladdnormallink{\protect {\tt{http://www.phys.lsu.edu/faculty/pullin}}}
{http://www.phys.lsu.edu/faculty/pullin}\\
\hfill ISSN: 1527-3431
\begin{rawhtml}
<P>
<BR><HR><P>
\end{rawhtml}
\end{flushleft}
\pagebreak
\section*{Editorial}

Not much to report here. This newsletter is a bit late, since I wanted
to wait till LIGO could report on its first science run. I want to 
encourage the readership to suggest topics for articles in MOG. In the
last few issues  articles were solicited  by myself. This 
is not good for keeping the newsletter balanced. Either contact the
relevant correspondent or me directly.

The next newsletter is due September 1st. All issues are available in the
WWW:\\\htmladdnormallink{\protect {\tt{http://www.phys.lsu.edu/mog}}}
{http://www.phys.lsu.edu/mog}\\ 

The newsletter is  available for
Palm Pilots, Palm PC's and web-enabled cell phones as an
Avantgo channel. Check out 
\htmladdnormallink{\protect {\tt{http://www.avantgo.com}}}
{http://www.avantgo.com} under technology$\rightarrow$science.

A hardcopy of the newsletter is
distributed free of charge to the members of the APS
Topical Group on Gravitation upon request (the default distribution form is
via the web) to the secretary of the Topical Group. 
It is considered a lack of etiquette to
ask me to mail you hard copies of the newsletter unless you have
exhausted all your resources to get your copy otherwise.
\par
If you have comments/questions/complaints about the newsletter email
me. Have fun.
\bigbreak
   
\hfill Jorge Pullin\vspace{-0.8cm}
\parskip=0pt
\section*{Correspondents of Matters of Gravity}
\begin{itemize}
\setlength{\itemsep}{-5pt}
\setlength{\parsep}{0pt}
\item John Friedman and Kip Thorne: Relativistic Astrophysics,
\item Raymond Laflamme: Quantum Cosmology and Related Topics
\item Gary Horowitz: Interface with Mathematical High Energy Physics and
String Theory
\item Beverly Berger: News from NSF
\item Richard Matzner: Numerical Relativity
\item Abhay Ashtekar and Ted Newman: Mathematical Relativity
\item Bernie Schutz: News From Europe
\item Lee Smolin: Quantum Gravity
\item Cliff Will: Confrontation of Theory with Experiment
\item Peter Bender: Space Experiments
\item Riley Newman: Laboratory Experiments
\item Warren Johnson: Resonant Mass Gravitational Wave Detectors
\item Gary Sanders: LIGO Project
\item Peter Saulson: former editor, correspondent at large.
\end{itemize}
\section*{Topical Group in Gravitation (GGR) Authorities}
Chair: Richard Price; Chair-Elect: John Friedman; Vice-Chair: Jim Isenberg;
Secretary-Treasurer: Patrick Brady; Past Chair: Bob Wald; Members at Large:
Gary Horowitz, Eric Adelberger, Ted Jacobson, Jennie Traschen, 
{\'E}anna Flanagan, Gabriela Gonz\'alez.
\vfill
\pagebreak

\parskip=10pt
\section*{\centerline {
GGR Activities}}
\addtocontents{toc}{\protect\medskip}
\addtocontents{toc}{\bf GGR News:}
\addcontentsline{toc}{subsubsection}{\it  
GGR Activities, by Richard Price}
\begin{center}
Richard Price, University of Utah (for the Exec. Committee)
\htmladdnormallink{rprice@physics.utah.edu}
{mailto:rprice@physics.utah.edu}
\end{center}

Our Topical Group in Gravitation continues to be a success when
measured by its size, as well as by any other metric.  We have 1.4\% of
the total APS membership.  Of the 9 APS Topical Groups, GGR is number
3 in size. It is ironic that our membership has grown, but that we
recently slipped from our number 2 position due to a jump in
membership of the magnetism topical group. (It is tempting to make
puns about many body phenomena here.)  The topical group in
Statistical and Nonlinear Physics remains in first place with 1.6\% of
APS membership.  The larger subunits of the APS are Divisions, with a
minimum of 3\% of APS membership.  Our GGR members come mostly from two
APS Divisions: the Division of Astrophysics and the Division of
Particles \& Fields.

A new GGR activity this past year is the GGR sponsorship of student
speaker awards at the regional gravity conferences. A new activity,
long in the development, is the presentation of the Einstein awards,
at a special Awards session (b6, Saturday 14:30) of the April APS
meeting in Philadelphia.  As always, a business meeting of our topical
group will be held during the April meeting.

Awaiting us further in the future is the international celebration of
the World Year of Physics in 2005. Judy Franz, Executive Director of
the APS, is the head of US participation in the event.  The
inspiration for all this is the centenary of Einstein's 1905 papers,
so GGR --- as the APS unit most closely associated with Einstein ---
should play some role. The Executive Committee has been discussing
ways in which this is to be done, but would benefit from suggestions.

\vfill
\eject

\section*{\centerline {
We hear that...}}
\addcontentsline{toc}{subsubsection}{\it  
We hear that... by Jorge Pullin}
\begin{center}
Jorge Pullin, Louisiana State University
\htmladdnormallink{pullin@phys.lsu.edu}
{mailto:pullin@phys.lsu.edu}
\end{center}

{\bf Stan Whitcomb, Lee Lindblom, Sam Finn, Ashok Das} 
were elected fellows of the
American Physical Society.

{\bf Vicky Kalogera} received a David and Lucile Packard Fellowship
and the A. J. Cannon Award of the American Astronomical Society and
the American Association of University Women.

{\bf Barry Barish} was elected to the National Academy of Sciences and
to the National Science Board.

{\bf Luis Lehner} was awarded a Sloan Fellowship.

The Prime Minister of the UK, Tony Blair, opened {\bf The Ogden Centre
for Fundamental Physics at the University of Durham} on Friday 18th
October 2002. The multi-million pound science complex will create a
world-leading centre of excellence in fundamental physics, combining
research into the building blocks of the universe and the large scale
structure of the universe, coupled with a mission to inspire a new
generation of young scientists.

Hearty congratulations!
\vfill
\eject

\section*{\centerline {
Institute of Physics Gravitational Physics Group}} 
\addcontentsline{toc}{subsubsection}{\it  
Institute of Physics Gravitational Physics Group, by Elizabeth Winstanley}
\begin{center}
Elizabeth Winstanley, University of Sheffield
\htmladdnormallink{E.Winstanley@sheffield.ac.uk}
{mailto:E.Winstanley@sheffield.ac.uk}
\end{center}

The Gravitational Physics Group of the Institute of Physics (UK) was formed
at the beginning of 2002.  The aim of the group is to form a
scientific and public focus for the gravitational physics community,
in particular, providing a forum for discussion between the theory and
experimental researchers and with UK funding bodies.

Prof. Mike Cruise (Birmingham) currently chairs the group and
Prof. Ray d'Inverno (Southampton) is the secretary.  Details of the
other members of the committee can be found on the group's web-site
\htmladdnormallink{http://groups.iop.org/GP/}{http://groups.iop.org/GP/}. 
The committee includes theorists and experimentalists with a wide
range of research interests and also a representative from industry.
The committee representative from BAE Systems, provides a link between
industry and academia, enabling liaison for exploring and directing
industry supported university programmes in this and related areas of
research.

A principal concern of the Group is the funding of Gravitational
Physics in the UK where, partly for historic reasons, funding has been
inadequate, especially on the theoretical side, since it has often
appeared to have fallen between the responsibilities of the two main
funding agencies PPARC and EPSRC. The Group plans to carry out a
survey of the field in the UK to find out some basic information about
its size and composition, prior to setting up a meeting with the two
funding agencies to explore ways of increasing the funding base.

The group's first scientific meeting, held in London in February 2002,
discussed the interplay between theory and experiment in gravitational
physics, with talks by Dr. B.S. Sathyaprakash (Cardiff) on
``Gravitational physics: the interaction of theory and experiment'',
and Prof. Gary Gibbons (Cambridge) on ``Gravitational Physics: General
Relativity and its role in fundamental physics'', as well as a
presentation on PPARC funding for Gravitational Physics research by
Prof. Richard Wade (PPARC).  Subsequently, the group hosted a meeting
on ``Brane World Gravity'', in November 2002 (see separate report),
and a joint meeting with the Royal Astronomical Society on
``Gravitational Wave Astronomy'', in February 2003.

On the industrial side, BAE Systems chaired a meeting in November
2002, attended by Government agencies (MoD, Qinetiq, Dstl), Rolls
Royce and several UK universities on research into applications of
gravitational physics for advanced propulsion systems.

The group also supports the annual BritGrav meetings.  The third
British Gravity Meeting is to be hosted by Lancaster University
Physics Department, and will take place on 12 - 14 September, 2003.
For further details see the web-site: 
\htmladdnormallink
{http://www.lancs.ac.uk/depts/physics/conf/britgrav/}
{http://www.lancs.ac.uk/depts/physics/conf/britgrav/}.

\vfill
\eject
\section*{\centerline {
Center for Gravitational Wave Astronomy}\\ 
\centerline{a new NASA University Research Center}}
\addcontentsline{toc}{subsubsection}{\it  
Center for Gravitational Wave Astronomy, by Mario D\'{\i}az}
\begin{center}
Mario D\'{\i}az, University of Texas at Brownsville
\htmladdnormallink{mdiaz@utb.edu}
{mailto:mdiaz@utb.edu}
\end{center}

On January 1st 2003, the National Aeronautic and Space Administration (NASA)
created at The University of Texas at Brownsville (UTB) the Center
for Gravitational Wave Astronomy (CGWA) as part of its University
Research Center (URC) program. As described by NASA, the University
Research Centers program is designed to achieve a
broad-based, competitive aerospace research capability among the
nation's Minority Serving Institutions (MSI) of Higher Education.
These centers will foster new aerospace science and technology 
concepts; expand the nation's base for aerospace research and 
development; develop mechanisms for increased participation by 
faculty and students of MSI in mainstream research; and increase the
production of socially- and economically-disadvantaged students 
(who are U.S.\ citizens and who have historically been
underrepresented) with advanced degrees in NASA-related fields.

This particular center will develop excellence in research and 
education in areas related to the new astronomy which will become 
technically feasible within the next five to ten 
years---{\em gravitational wave astronomy}.

The CGWA will focus on three major research areas: 
gravitational wave data analysis, gravitational wave source modeling, 
and phenomenological
astrophysics of supermassive black holes. The proposed research is
relevant to the NASA Space Science Enterprise of charting the
evolution of the universe and understanding its galaxies, stars,
and their dynamics and evolution. In particular, we expect the
center to make important contributions to LISA, a joint NASA-ESA
mission with a projected launch date of 2011.  LISA consists
of three identical spacecraft located in an equilateral triangle
$5\times10^6$ km on a side in a heliocentric orbit.  The
spacecraft carry the optical components of a Michelson-Morley
interferometer, which will measure the passing of gravitational
waves of astrophysical origin in the $10^{-1}$ to $10^{-4}$ Hz
band. NASA's recognition of the technological and
scientific opportunities presented by the LISA mission is
exemplified in their selection of the Disturbance Reduction System
technology as the Space Technology 7 project for the New
Millennium Program.

As already evidenced by the LIGO project, the success of LISA
does not depend only on the expertise of the experimental
scientists and engineers who design and implement it, but also 
on the collaboration of a highly integrated group of
scientists in astrophysics, source modeling, and data analysis.
Data analysts rely on source modelers and astrophysicists to
predict features of gravitational wave signals that allow
them to be extracted from instrumental noise. Source modelers rely
on astrophysicists and data analysts to guide them in representing
the sources that are most likely to be observed. Astrophysicists
use source modeling and signals extracted by data analysts (or
the lack of such signals) to improve their understanding of the
astrophysics of the actual sources.

The proposed CGWA will represent research expertise in all three
of these theoretical disciplines, with a focus on LISA research.
The core of the center's research personnel will be formed by the
UTB Relativity Group (UTBRG). Currently, the UTBRG consists of
five faculty (Warren Anderson, Manuela Campanelli, Mario
D\'{\i}az, Carlos Lousto and Joe Romano), three post-docs and
several undergraduate and graduate students. Since its creation,
just six years ago, the UTBRG has developed expertise in
gravitational wave source modeling and data analysis. Center
funding will allow existing research strengths to be augmented by
expanding  the area of phenomenological astrophysics---especially
related to super-massive black holes---with the incorporation of
more scientists (at both the faculty and postdoctoral level).
The Center is advised at the scientific level by a
very distinguished Board of Advisors: 
Peter Bender (JILA-University of Colorado), Tom Prince (Caltech), 
Jorge Pullin (LSU),
Douglas Richstone (University of Michigan), and Bernard Schutz (AEI).

The strong research programs at the center will help it become 
a national and international hub for gravitational wave
scientists. It will complement a network of recently created
institutions/centers devoted to gravitational wave physics. The
UTBRG already has strong ties to the NSF Physics Frontier Center
for Gravitational Wave Physics recently created at Penn State,
to the LIGO Scientific Collaboration (which provides
scientific support to ground-based gravitational wave detectors),
and to the recently formed source modeling group at NASA's 
Goddard Space
Flight Center. These affiliations will provide additional research
avenues to center students and faculty. The center will have a
very active visitors program. All center activities are open to
the broad scientific community, whose participation will be
supported through this program.

In addition, the
center will develop one of its major thrusts in student support
at the advanced undergraduate and graduate levels.
Starting next year, the center will organize, on an
annual basis, an advanced undergraduate/beginning graduate
summer school in gravitational wave physics.

At the end of this year, the CGWA will hold its inaugurational
meeting. For more information about this event and the
opportunities provided by the center, please visit the web site
(presently under construction) at 
\htmladdnormallink{http://cgwa.phys.utb.edu}
{http://cgwa.phys.utb.edu}.

\vfill
\eject

\section*{\centerline {
LIGO's First Preliminary Science Results}}
\addtocontents{toc}{\protect\medskip}
\addtocontents{toc}{\bf Research Briefs:}
\addcontentsline{toc}{subsubsection}{\it  
LIGO's first preliminary science run, by Gary Sanders}
\begin{center}
Gary Sanders, LIGO-Caltech, on behalf of the LIGO Scientific Collaboration
\htmladdnormallink{sanders@ligo.caltech.edu}
{mailto:sanders@ligo.caltech.edu}
\end{center}

In the previous report in MOG [1], I described the completion of the first
LIGO science run, S1. The strain spectral sensitivity of LIGO in that
run, and the duration of the run, is described in that article. S1 was
the first in a series of progressively more sensitive science runs,
interleaved with interferometer commissioning and improvement
periods. S1 has now yielded preliminary science results. These are
being presented, as this article is written, at the AAAS meeting in
Denver [2]. The talks from this symposium will be posted shortly on the
LIGO Laboratory website [3], and the LSC website [4]. This article follows
closely the content of Albert Lazzarini's talk in that
symposium. Final results from S1 are in preparation and should appear
as preprints in the next several weeks with formal publication after
the March LSC meeting.

The results below mark the first from LIGO, a milestone in the
program. The physics impact of these results is not very significant,
but the lessons learned about the interferometer data streams and the
analysis algorithms lay the foundation for more meaningful future work
as LIGO's sensitivity increases and as the science runs extend over
longer periods. In that sense, the analysis methods that will be
published are at least as significant as the results themselves.

The GEO-600 interferometer also operated during S1, as we previously
reported, and results are quoted below from the joint GEO-LIGO
effort. Coincident running with the TAMA-300 interferometer is the
subject of analysis by a separate working group and is not reported
here.

For the analysis of this data, the LIGO Scientific Collaboration has
divided itself into four Upper Limits Working Groups. The terminology
clearly reflects the expectation that the early running is most useful
to set upper limits on the various gravitational wave fluxes and/or
source populations. These working groups are focused on burst,
inspiral, periodic and stochastic sources.

The data analysis techniques adapt to the source
characteristics. Deterministic signals from inspiral and periodic
sources can be parameterized by amplitude and frequency
evolution. Template sets can be selected to cover the parameter space
covered by the data. Statistical signals, such as stochastic
gravitational wave backgrounds, are sought by cross-correlating pairs
of interferometers, seeking correlations and statistical
variations. Unmodeled signals such as supernovae, gamma ray bursts or
entirely new transient sources cannot be addressed by parametric
techniques. For these, very basic data trends are sought such as power
excesses in the frequency-time domain, or notable amplitude changes in
time. In all of these studies, use is made of the coincidence between
multiple detectors, a most powerful filter.

The LIGO Burst Sources Working Group has taken on one of the most
challenging search topics. In the absence of a waveform model, they
have searched for frequency vs. time domain power excesses and time
varying amplitude departures, setting a bound on unmodeled bursts of
rate vs. strength. The analysis serves as a prototype for any searches
for discrete events. Diagnostic triggers from the detectors indicating
instrumental or environmental data artifacts confront event triggers
from the interferometers. Requiring three-interferometer coincidence
further filters event triggers that are not vetoed by this
process. Bursts of peak strain amplitude above a given rate have been
excluded in this search as shown in Figure 1. The displayed limit is
not yet the best achieved as resonant mass detectors have published
superior limits [5,6].

\begin{figure}[h]
\centerline{\psfig{figure=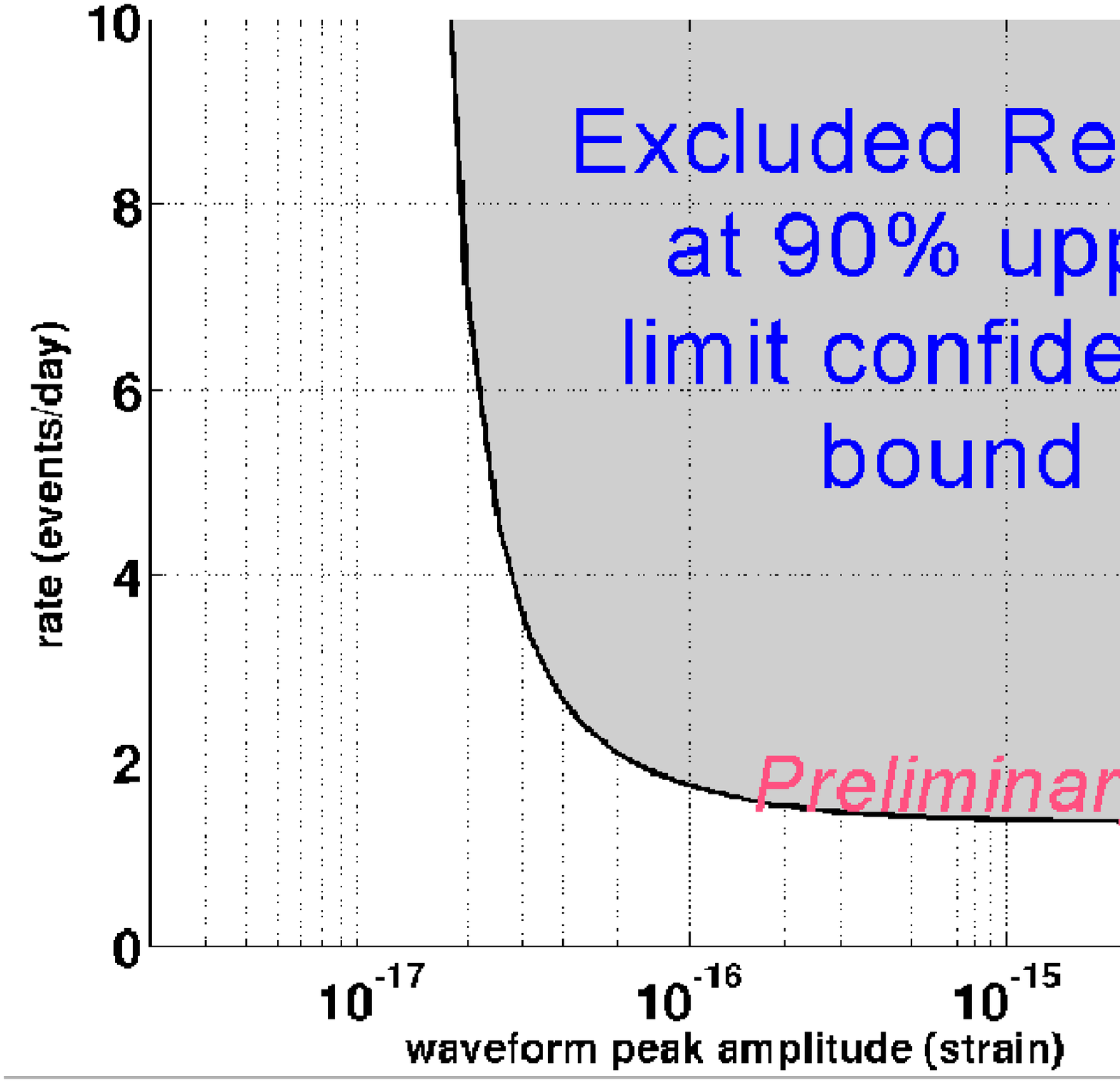,height=4in}}
\end{figure}

The Inspiral Sources Working Group is searching for 1- 3 solar mass
neutron star binaries, black hole binaries heavier than 3 solar
masses, and MACHO binaries in the 0.5 - 1 solar mass range. They have
completed the neutron star search at this time. The analysis employs
template based matched filtering. During the S1 run, the LIGO
interferometers were individually sensitive to 1.4 solar mass binaries
out to 38 kpc to 210 kpc, for optimal orientation and signal to noise
ration (SNR) of 8. With appropriate treatment of the expected
population in the Milky Way galaxy, and the Large and Small Magellanic
Clouds, and using triggers from the Hanford and Livingston 4 km
interferometers, the resulting limit is:

$$ {\rm R90}\,\%\,{\rm (Milky Way)} <  1.7 \times 10^2 /{\rm yr}$$

The best previously published observational limit was obtained using
data from the 40 Meter Interferometer at Caltech [7] ($
{\rm R90}\,\%\,{\rm (Milky Way)} < 
4.4 \times 10^3 /{\rm yr}$).

The Periodic Sources Working Group is searching for gravitational
waves radiated by periodic sources such as rotating prolate neutron
stars with ellipticity in the range $10^{-3} - 10^{-4}$. These would
be of best interest in a LIGO search if all of the observed spin-down
were attributable to gravitational wave emissions. The search can be
carried out in the frequency domain by cross-correlating the data
stream with templates, seeking power correlations. A complementary
approach uses a time domain search, removing motion of the Earth, and
comparing the result with what would be expected from noise in the
absence of a signal. The analyses are still in progress. However, no
signals are seen. Preliminary limits, with 95\% confidence, on periodic
sources are set on maximum strain amplitudes individually for the GEO
interferometer and the three LIGO interferometers ranging from $3 \times
10^{-21}$ to $2 \times 10^{-22}$, the latter limit set by the Livingston 4 km
interferometer.

The Stochastic Sources Working Group is searching for these remnant
signals by cross-correlating interferometer outputs in pairs, using
the long baseline Livingston-Hanford 4 km interferometers and the
short baseline co-located Hanford 2 km and 4 km interferometers. The
full initial LIGO science run should be able to reach sensitivity of
$\sim \Omega_{GW}< 10^{-5}$, comparable to upper limits inferred from Big Bang
nucleosynthesis [8] . The best detector-based limit is GW 907 Hz from
resonant mass detector results [9]. Based upon analysis of only 7.5 hours
of data from the S1 run, the Stochastic Sources Working Group
estimates the limits from the full S1 run as shown in Table 1.

\begin{table}
\begin{tabular}{|c|c|c|c|}
\hline
Interferometer & Measurement & Extrapolated Upper Limit & ${\rm T_{obs}}$\cr
Pair & bandwidth & for S1 (by scaling 7.5 hrs to 150 hrs) & \cr
\hline
H2km - H4km & 40Hz 
$<$ f $<$ 300 Hz & $\Omega_{\rm GW} < 5$ (90\% C.L.) & 150 hr\cr
\hline
H4km - L4km & 40Hz 
$<$ f $<$ 314 Hz & $\Omega_{\rm GW} < 70$ (90\% C.L.) & 100 hr\cr
\hline
H2km - L4km & 40Hz 
$<$ f $<$ 314 Hz & $\Omega_{\rm GW} < 500$ (90\% C.L.) & 100 hr\cr
\hline
\end{tabular}
\end{table}

This is only an estimate of the S1 result, as stated. Actual results
from the full data set will be represented in preprints to be released
prior to the LSC meeting in March.

\begin{figure}[h]
\centerline{\psfig{figure=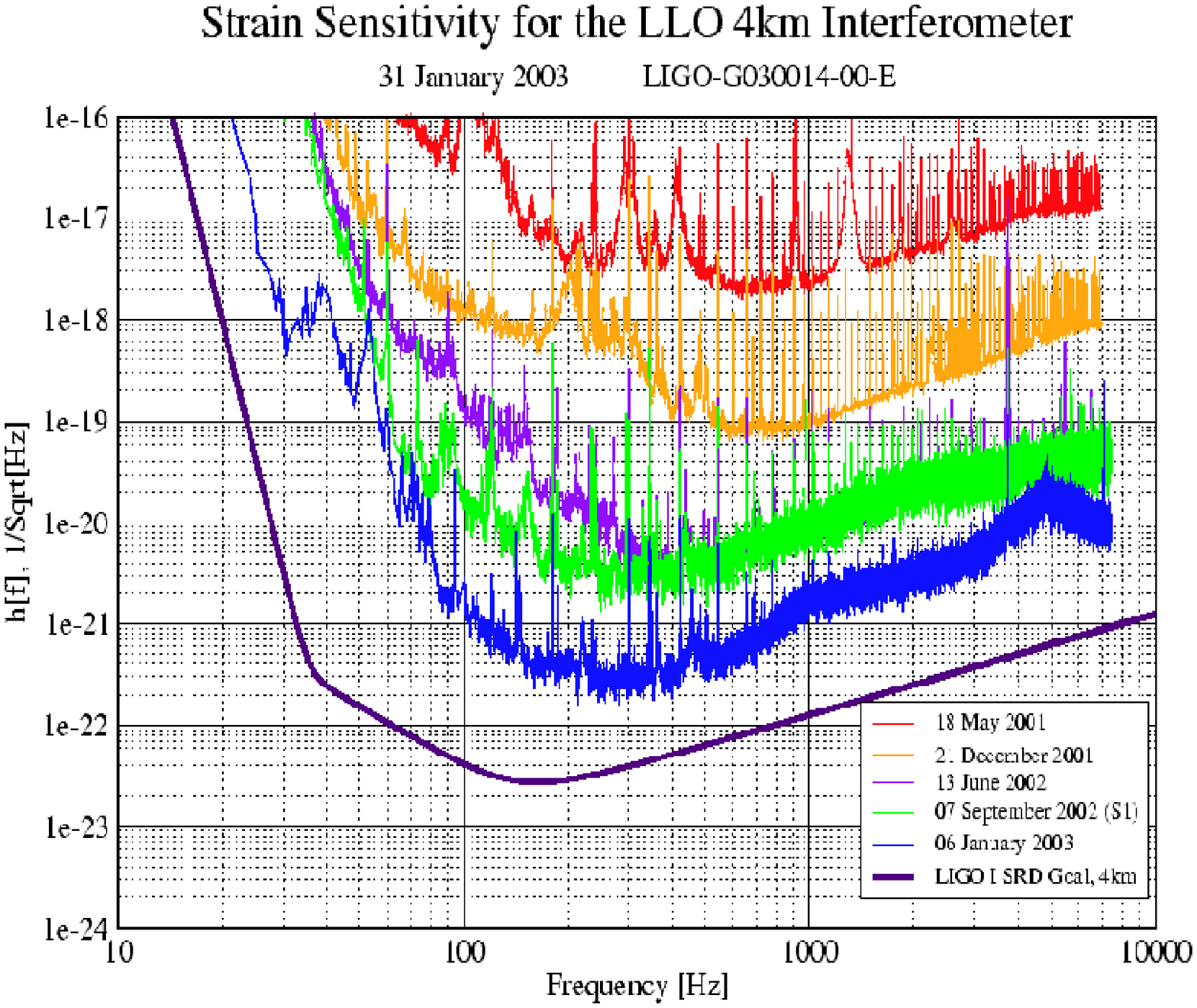,height=4.5in}}
\end{figure}

As this S1 analysis matures, LIGO has already begun the S2 run, on
February 14. An 8-week run is planned, four times longer than the S1
run. Most important, LIGO sensitivity is more than an order of
magnitude better now, across the sensitive spectrum. Figure 2 displays
the progression in sensitivity for the Livingston 4 km interferometer,
illustrating the improvement since S1.

The S2 run will involve coincident running with the LSC partner
GEO-600 instrument, and with the TAMA-300 detector, following the
signing of a new LIGO-TAMA Memorandum of Understanding. 2003 promises
to be a propitious year for ground-based gravitational wave
interferometry!

LIGO is funded by the US National Science Foundation under Cooperative
Agreement PHY-0107417. This work is a collaborative effort of the
Laser Interferometer Gravitational-wave Observatory and the
institutions of the LIGO Scientific Collaboration. LIGO T030021-00-M.

More information about LIGO can be found at: 
\htmladdnormallink{http://www.ligo.caltech.edu} 
{http://www.ligo.caltech.edu}.

{\bf References:}

[1] Previous Mog article:
\htmladdnormallink
{http://www.phys.lsu.edu/mog/mog20/node10.html}
{http://www.phys.lsu.edu/mog/mog20/node10.html}

[2] AAAS meeting
\htmladdnormallink
{http://php.aaas.org/meetings/MPE\_01.php?detail=1198}
{http://php.aaas.org/meetings/MPE\_01.php?detail=1198}

[3] LIGO website
\htmladdnormallink
{http://www.ligo.caltech.edu}
{http://www.ligo.caltech.edu}

[4] LSC website
{http://www.ligo.org}
{http://www.ligo.org}

[5] G. A. Prodi et al.  Int. J. Mod. Phys. D9 (2000) 237

[6] P. Astone et al.   Class.Quant.Grav. 19 (2002) 5449

[7] B.  Allen et al., Phys.Rev.Lett. 83 (1999) 1498

[8]  E. Kolb, M. Turner ``The early universe'', Addison Wesley  (1990).

[9]  P. Astone et al., Astron. Astrophys. 351, 811 (1999).

\vfill
\eject

\section*{\centerline {
Quantization of area: the plot thickens}}
\addcontentsline{toc}{subsubsection}{\it  
Quantization of area: the plot thickens, by John Baez}
\begin{center}
John Baez, University of California at Riverside
\htmladdnormallink{baez@math.ucr.edu}
{mailto:baez@math.ucr.edu}
\end{center}

One of the key predictions of loop quantum gravity is that the area of a
surface can only take on a discrete spectrum of values.  In particular,
there is a smallest nonzero area that a surface can have.  We can call
this the `quantum of area', so long as we bear in mind that not all
areas are integer multiples of this one --- at least, not in the most
popular version of the theory.

So far, calculations working strictly within the framework of loop
quantum gravity have been unable to determine the quantum of area.  But
now, thanks to work of Olaf Dreyer [1] and Lubo\v s Motl
[2], two very different methods of calculating the quantum of
area have been shown to give the same answer: $4 \ln 3$ times the Planck
area.  Both methods use semiclassical ideas from outside loop quantum
gravity.  The first uses Hawking's formula for the entropy of a black
hole, while the second uses a formula for the frequencies of highly
damped vibrational modes of a classical black hole.  It is still 
completely mysterious why they give the
same answer.  It could be a misleading coincidence, or it could be an
important clue.  In any event, the story is well worth telling.

The importance of {\it area} in quantum gravity has been obvious
ever since the early days of black hole thermodynamics.  In
1973, Bekenstein [3] argued that the entropy of a 
black hole was proportional to its area.   By 1975, Hawking 
[4] was able to determine the constant of proportionality,
arriving at the famous formula
$$
S = A/4
$$
in units where $\hbar = c = G = k = 1$.
Understanding this formula more deeply has been a challenge ever since.

Things took a new turn around 1995, when Rovelli and Smolin [5]
showed that in loop quantum gravity, area is quantized.
The geometry of space is described using `spin networks', which are 
roughly graphs with edges labeled by spins:

\vskip 2em
\centerline{\epsfysize=2.0in\epsfbox{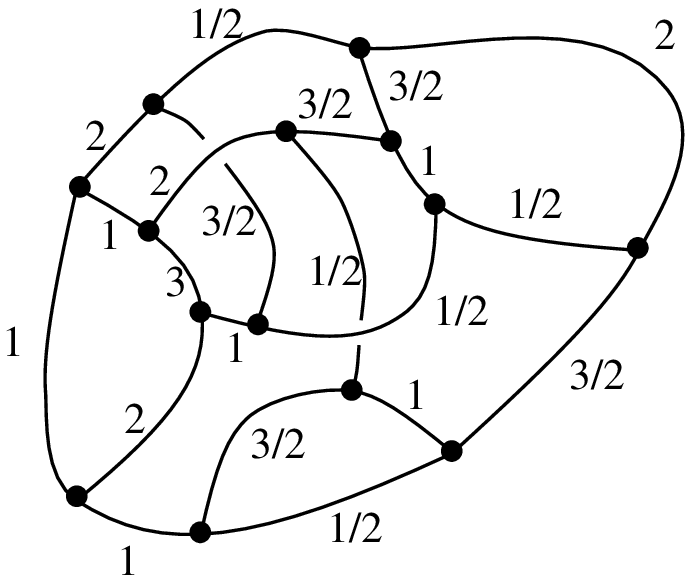}} \medskip

\noindent
Any surface gets its area from spin network edges
that puncture it, and an edge labeled by the spin $j$ contributes
an area of $8 \pi \gamma \sqrt{j(j+1)}$, where
$\gamma$ is a dimensionless quantity called the Barbero--Immirzi
parameter [6,7].  

Given this, it was tempting to attribute the entropy of a black hole to
microstates of its event horizon, and to describe these in 
terms of spin network edges puncturing the horizon.  After some
pioneering work by Rovelli and Smolin, Krasnov [6]
noticed that the horizon of a nonrotating black hole could be described
using a field theory called Chern-Simons theory.  He began working with
Ashtekar, Corichi and myself on using this to compute
the entropy of such a black hole.

By 1997 we felt we were getting somewhere, and we came out with a short
paper outlining our approach [7].  While the details are
technical [8], the final calculation is easy to describe.  The
geometry of the event horizon is described not only by a list of nonzero spins
$j_i$ label ling the spin network edges that puncture the horizon, but
also by a list of numbers $m_i$ which can range from $-j_i$ to $j_i$ in
integer steps.  The intrinsic geometry of the horizon is flat except at
the punctures, and the numbers $m_i$ describes the angle deficit at each
puncture.  To count the total number of microstates of a black hole of
area near $A$, we must therefore count all lists $j_i$, $m_i$ for which
$$ 
A \cong \sum_i 8 \pi \gamma \sqrt{j_i(j_i+1)}.  
$$

This is a nice little math problem.  It turns out that for a large black hole,
the whopping majority of all microstates come from taking all the spins
to be as small as possible.   So, we
can just count the microstates where all the spins $j_i$ equal $1/2$.  
If there are $n$ punctures, this gives
$$
A \cong 4 \pi \sqrt{3}\gamma n.
$$
In a state like this, each number $m_i$ can take just two values at each 
puncture.  Thus if there are $n$ punctures, there are $2^n$
microstates, and the black hole entropy is 
$$
S \cong \ln (2^n) \cong {\ln 2 \over 4 \pi \sqrt{3}\, \gamma } A.
$$  

In short, we see that entropy is indeed proportional to area, at least 
for large black holes.  However, we only get Hawking's formula
$S = A/4$ if we take the Barbero--Immirzi parameter to be
$$
       \gamma =  {\ln 2 \over \pi \sqrt{3}}.
$$
On the one hand this is good: it's a way to determine the
Barbero--Immirzi parameter, and thus the quantum of area, which works
out to 
$$    8 \pi \gamma \sqrt{\textstyle{1\over 2}(\textstyle{1\over 2}+1)} = 
4 \ln 2 .
$$
This makes for a pretty picture in which almost all
the spin network edges puncturing the event horizon carry one quantum of
area and one qubit of information, as in Wheeler's `it from bit' scenario
[9].  One can also check that the same value of $\gamma$
works for electrically charged black holes and black holes coupled to a
dilaton field.  On the other hand, it seems annoying that we can only
determine the quantum of area with the help of Hawking's semiclassical
calculation.  The strange value of $\gamma$ might also make us
suspicious of this whole approach.

Meanwhile, as far back as 1974, Bekenstein [10] had argued
that Schwarzschild black holes should have a discrete spectrum of evenly
spaced areas.  While this law does not hold in the loop quantum gravity
description of black holes, it has some of the same consequences.
For example, in 1986 Mukhanov [11] noted that with a law
of this sort, the formula $S = A/4$ can only hold exactly if the $n$th
area eigenstate has degeneracy $k^n$ and the spacing between area
eigenstates is $4 \ln k$ for some number $k = 2,3,4,$....  He also
gave a philosophical argument that the value $k = 2$ is preferred, since
then the states in the $n$th energy level can be described using $n$
qubits.

Many researchers have continued this line of thought in different ways,
but in 1995, Hod [14] gave an remarkable argument in favor
of $k = 3$.
His idea was to determine the quantum of area by looking at the
vibrational modes of a {\it classical} black hole!  Hod argues that
if classically a system can undergo periodic motion at some frequency
$\omega$, then in the quantum theory it can emit or absorb quanta of
radiation with the corresponding energy.  But the energy of a
Schwarzschild black hole is just its mass, and this is related to the
area of its event horizon by
$$
A = 16 \pi M^2,
$$
so when a black hole absorbs one quantum of radiation its area
should change by
$$
\Delta A = 32 \pi M \Delta M 
         = 32 \pi M \omega.
$$
And now for the miracle!  A nonrotating black hole will exhibit damped
oscillations when you perturb it momentarily in any way, and there are
different vibrational modes called quasinormal modes, each with its 
own characteristic frequency and damping.  In 1993, Nollert 
[15]  used computer calculations to 
show that in the limit of large damping, the frequency of these modes
approaches a specific number depending only on the mass of the black
hole:
$$
\omega \cong 0.04371235 / M.
$$
Plugging this into the previous formula, Hod obtained the
quantum of area 
$$
\Delta A = 4.394444 
$$
and noticed that this was extremely close to $4 \ln 3 =  4.394449$.
On the basis of this, he daringly concluded that $k = 3$.

Our story now catches up with recent developments.  
In November 2002, Dreyer [1] found an ingenious way to 
reconcile Hod's result with the loop quantum gravity calculation.  The
calculation due to Ashtekar {\it et al} used a version of
loop quantum gravity where the gauge group is ${\rm SU}(2)$.  This is why so
many formulas resemble those familiar from the quantum mechanics of
angular momentum, and this is why the smallest nonzero area comes from a
spin network edge labeled by the smallest nonzero spin: $j = 1/2$.  But
there is also a version of loop quantum gravity with gauge group
${\rm SO}(3)$, in which the smallest nonzero spin is $j = 1$.  
Dreyer observed out that if we repeat the black hole
entropy calculation using this ${\rm SO}(3)$ theory, we get a quantum of area
that matches Hod's result!  One can easily
check this by redoing the calculation sketched earlier, replacing
$j = 1/2$ by $j = 1$.  One finds a new value of the Immirzi parameter:
$$
\gamma =   {\ln 3 \over 2 \pi \sqrt{2}},
$$
and obtains $4 \ln 3$ as the new quantum of area.  But ultimately,
all that really matters is that when $j = 1$ there are 3 spin states
instead of 2.  Thus each quantum of area carries
a `trit' of information instead of a bit, which is why Dreyer obtains 
$k = 3$.

With the appearance of Dreyer's paper, the suspense became almost
unbearable.  After all, Hod's observation relied on numerical
calculations, so the very next digit of his number might fail to match
that of $4 \ln 3$.  Luckily, in December 2002, Motl [2]
showed that the match is exact!  He used an ingenious analysis
of Nollert's continued fraction expansion for the asymptotic
frequencies of quasinormal modes.  

While exciting, these developments raise even more questions than they
answer.  Why should ${\rm SO}(3)$ loop quantum gravity be the right
theory to use?  After all, it seems impossible to couple spin-1/2
particles to this version of the theory.  Corichi has sketched
a way out of this problem [16], but much work remains
to see whether his proposal is feasible.  Can we turn Hod's argument
from a heuristic into something a bit more rigorous?  He cites
Bohr's correspondence principle in this form: ``transition
frequencies at large quantum numbers should equal classical oscillation
frequencies.''  However, this differs significantly from the idea
behind Bohr--Sommerfeld quantization, and it is also unclear why we
should apply it only to the {\it asymptotic} frequencies of highly
damped quasinormal modes.  
Can the mysterious agreement between ${\rm SO(3)}$ loop quantum
gravity and Hod's calculation be extended to rotating black holes? 
Here a new paper by Hod makes some interesting progress [17].
Can it be extended to black holes in higher dimensions?  Here Motl's 
new work with Neitzke gives some enigmatic clues [17].  Stay 
tuned for further developments.

{\bf References:}

[1] O.\ Dreyer, \htmladdnormallink{gr-qc/0211076}
{http://arXiv.org/abs/gr-qc/0211076}.
\parskip=0pt
[2] L.\ Motl, \htmladdnormallink{gr-qc/0212096}
{http://arXiv.org/abs/gr-qc/0212096}.

[3] J.\ Bekenstein, {\sl Phys.\ Rev.\ }{\bf D7} 
(1973), 2333--2346.

[4] S.\ Hawking, {\sl Commun.\ Math.\ Phys.\ } {\bf 43} (1975), 199.

[5] C.\ Rovelli and L.\ Smolin, 
{\sl Nucl.\ Phys.\ } {\bf B442} (1995), 593.
Erratum: {\sl Nucl.\ Phys.\ }{\bf B456} (1995), 734. 
Also at \htmladdnormallink{gr-qc/9411005}
{http://arXiv.org/abs/gr-qc/9411005}. 

[6]
G.\ Immirzi, 
{\sl Nucl.\ Phys.\ Proc.\ Suppl.\ }{\bf 57} (1997), 65.  Also
at \htmladdnormallink{gr-qc/9701052}
{http://arXiv.org/abs/gr-qc/9701052}.

[7]
F.\ Barbero, {\sl Phys.\ Rev.\ }{\bf D51} (1995), 5507.
Also at \htmladdnormallink{gr-qc/9410014}
{http://arXiv.org/abs/gr-qc/9410014}.

[8]
K.\ Krasnov, {\sl Gen.\ Rel.\ Grav.\ }{\bf 30} (1998), 53.
Also at \htmladdnormallink{gr-qc/9605047}
{http://arXiv.org/abs/gr-qc/9605047}.

[9]
A.\ Ashtekar, J.\  Baez, A.\ Corichi and K.\ Krasnov,
 {\sl Phys.\ Rev.\ Lett.\ }{\bf 80}
(1998), 904.  Also at \htmladdnormallink{gr-qc/9710007}
{http://arXiv.org/abs/gr-qc/9710007}.  

[10]
A.\ Ashtekar, A.\ Corichi and K.\ Krasnov,
 {\sl Adv.\ Theor.\ 
Math.\ Phys.\ } {\bf 3} (2000), 418.  Also at \htmladdnormallink{gr-qc/9905089}
{http://arXiv.org/abs/gr-qc/9905089}.

A.\ Ashtekar, J.\ Baez, and K.\ Krasnov,
{\sl Adv.\ Theor.\ Math.\ Phys.\ }{\bf 4}
(2000), 1.  Also at \htmladdnormallink{gr-qc/0005126}
{http://arXiv.org/abs/gr-qc/0005126}.

[11] J.\ Wheeler, in {\sl Sakharov 
Memorial Lecture on Physics,} vol.\ 2, eds.\ L.\ Keldysh and V.\ 
Feinberg, Nova Science, New York, 1992.

[12]
J.\ Bekenstein, {\sl Lett.\ Nuovo Cimento} {\bf 11} (1974), 467.

[13] V.\ Mukhanov, {\sl 
JETP Lett.\ }{\bf 44} (1986), 63.

J.\ Bekenstein and V.\ Mukhanov, 
{\sl Phys.\ Lett.\ }{\bf B360} (1995), 7.  Also at \htmladdnormallink{gr-qc/9505012}
{http://arXiv.org/abs/gr-qc/9505012}.

[14] S.\ Hod,
{\sl Phys.\ Rev.\ Lett.\ }{\bf 81} (1998), 
4293. Also at \htmladdnormallink{gr-qc/9812002}
{http://arXiv.org/abs/gr-qc/9812002}.

S.\ Hod, {\sl Gen.\ Rel.\ Grav.\ } {\bf 31} (1999), 1639.
Also at \htmladdnormallink{gr-qc/0002002}
{http://arXiv.org/abs/gr-qc/0002002}.

[15]
H.-P.\ Nollert, 
{\sl Phys.\ Rev.\ }{\bf D47} (1993), 5253.

[16]
A.\ Corichi, \htmladdnormallink{gr-qc/0212126}
{http://arXiv.org/abs/gr-qc/0212126}.

[17] S.\ Hod, 
\htmladdnormallink{gr-qc/0301122}
{http://arXiv.org/abs/gr-qc/0301122}.

[18]
L.\ Motl and A.\ Neitzke, 
\htmladdnormallink{hep-th/0301173}
{http://arXiv.org/abs/gr-qc/0301173}.
\vfill
\eject

\parskip=5pt
\section*{\centerline {
Convergence (?) of G Measurements -- Mysteries Remain.}}
\addcontentsline{toc}{subsubsection}{\it  
Convergence of G Measurements -- Mysteries Remain, by Riley Newman}
\begin{center}
Riley Newman, University of California at Irvine
\htmladdnormallink{rdnewman@uci.edu}
{mailto:rdnewman@uci.edu}
\end{center}

In 1995 the German PTB G measurement group published a startling G
value [1] more than half a percent higher than the previously accepted
CODATA value.  The figure below indicates the most recent results of
groups reporting G values since then.

\begin{figure}[h]
\centerline{\psfig{figure=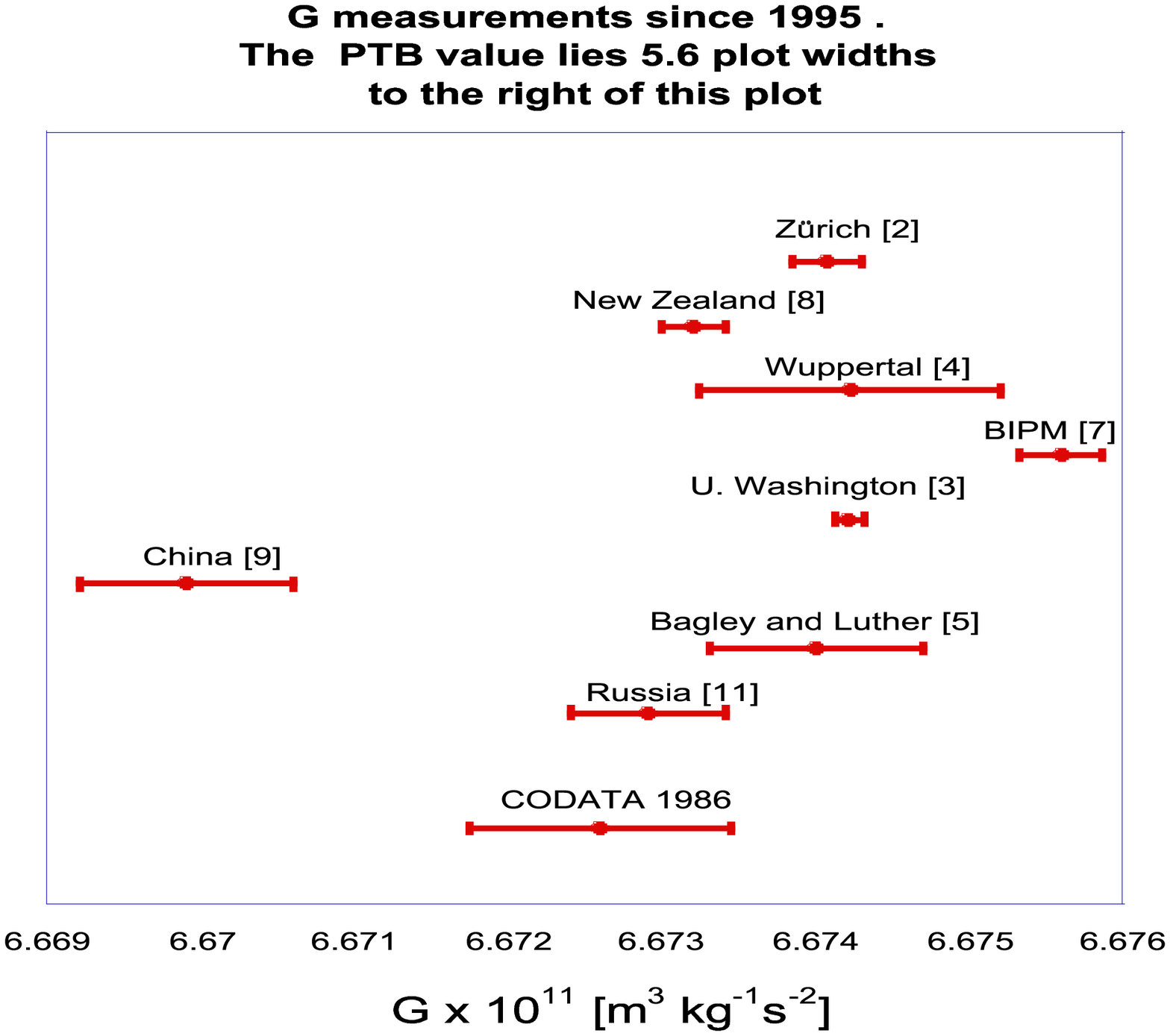,height=3in}}
\end{figure}

The recently published 33 ppm measurement [2] by the U. Z\"urich group
is nicely consistent with the 14 ppm measurement by the U. Washington
group [3] and with the 147 ppm measurement by the U. Wuppertal group
[4] and the 105 ppm measurement by Luther and Bagley [5].  These four
measurements used very different techniques: The Z\"urich group
measured the weight change of test masses caused by tanks of mercury
(this experiment is featured in last November's Physics Today [6].)
The Washington group measured the acceleration of test masses
suspended on a rotating platform servoed to match the test mass'
acceleration as source masses revolved about the system on a separate
platform.  The Wuppertal group measured the differential lateral
displacement produced by a source mass acting on a pair of separately
suspended test masses forming reflecting walls of a microwave cavity.
The measurement by Luther and Bagley at LANL used the ``time-of-swing"
technique of Luther and Towler's 1982 G measurement on which the 1986
CODATA value was based -- but this time deliberately used low-Q fibers
to reveal and then correct for the effect of fiber anelasticity.

The agreement of these measurements -- especially the U. Z\"urich and
U. Washington values -- is encouraging, but the figure shows
significant outliers.  Relative to the U. Washington G value, the BIPM
value [7] is 206 ppm or 4.9$\sigma$ {\bf higher }, while the most
recent New Zealand value [8] is 152 ppm or 4.6$\sigma$ {\bf lower}.
The value found by Jun Luo and collaborators in China [9] is also
lower, by 647 ppm or 6.1$\sigma$.

The BIPM measurement uses a torsion balance with two substantially
independent techniques, sharing the same mass and suspension
configuration.  One technique balances the gravitational forces on the
pendulum with a calibrated electrostatic force. The second technique
is that of Cavendish -- a measurement of static deflection, with
torsion constant calibration from the frequency of free oscillation.
The danger posed by fiber anelasticity is virtually eliminated by the
first technique (where the fiber does not twist significantly), and
highly suppressed in the second by using a thin strip suspension whose
restoring torque is largely gravitational in origin.  The two methods
give results in excellent agreement.  Further work on this measurement
is in progress.

The New Zealand group, like the U. Washington group, used a method in
which fiber anelasticity dangers are avoided by mounting a torsion
pendulum on a rotating platform servoed to match the pendulum's
angular acceleration so that the fiber twist is negligible.

The ``time-of-swing" measurement by Jun Luo and collaborators in China
minimized fiber anelasticity problems by using a very high Q
suspension.

Still not ready to report is my own G measurement group at UC Irvine,
which uses the ``time-of-swing" method with a cryogenic high-Q torsion
pendulum.  We are troubled by a timing glitch in some of our early
data, temperature control problems in more recent data, and most
disturbingly, evidence that the value of G we would obtain may depend
on the pendulum's oscillation amplitude.

The 1986 CODATA G value became suspect when Kuroda [10] pointed out
that fiber anelasticity would have biased Luther's ``time-of-swing"
measurement on which it was based toward a {\bf higher} G value.  The
five most recent measurements of G, uppermost in the figure above, all
have designs which virtually eliminate anelasticity dangers.
Interestingly, the 1986 CODATA value is actually {\bf lower } than the
value reported by all five of these measurements, as well as the PTB
measurement.

It seems clear that the PTB G value cannot be correct.  The PTB
measurement balanced gravitational forces against DC electrostatic
forces whose calibration involved the measurement of capacitances with
1 KHz AC voltages.  A danger which the experimenters recognized is
that the capacitances could be frequency dependent, and indeed the
BIPM paper [7] suggests that a grounded lossy dielectric in the fringe
field of the capacitive electrodes could cause such dependence.  But
the PTB workers did check for frequency dependence over a frequency
range 100 Hz to 1 KHz and found negligible effect -- it would seem
surprising if a 0.6\% capacitance calibration error that could account
for the anomalous G value escaped detection.  So mysteries remain, and
measured G values have not converged satisfactorily, although the
excellent agreement of the Z\"urich and U. Washington G values is
encouraging.\\

{\bf References:}

\parskip=0pt
[1] W. Michaelis, H. Haars and R. Augustin, Metrologia {\bf12}, 267 (1995/96)

[2] St. Schlamminger, E. Holzschuh, and W. K\"undig, Phys. Rev. Lett
{\bf89}, 161102 (2002)

[3] Jens H. Gundlach and Stephen M. Merkowitz, Phys. Rev. Lett
{\bf85}, 2869 (2000)

[4] Hinrich Meyer et al, reported by Jim Faller at the CPEM2002
metrology conference, Ottawa (June 2002)

[5] Charles H. Bagley and Gabriel G. Luther, Phys. Rev. Lett {\bf78},
3047 (1997)

[6] Physics Today {\bf55}, 19 (November, 2002)

[7] T.J. Quinn, C.C. Speake, S.J. Richman, R.S. Davis, and A. Picard,
Phys. Rev. Lett {\bf87}, 111101 (2001)

[8] T. Armstrong, reported at the CPEM2002 metrology conference,
Ottawa (June 2002)

[9] Jun Luo et al, Phys. Rev {\bf D59}, 042001 (1998)

[10] K. Kuroda, Phys. Rev. Lett. {\bf75}, 2796 (1995)

[11] O. Karagioz and V. Izmailov, Meas. Tech. {\bf39}, 979 (1996)

\parskip=5pt
\vfill
\eject

\section*{\centerline {
Brane world gravity}}
\addtocontents{toc}{\protect\medskip}
\addtocontents{toc}{\bf Conference reports:}
\addcontentsline{toc}{subsubsection}{\it  
Brane world gravity, by Elizabeth Winstanley}
\begin{center}
Elizabeth Winstanley, University of Sheffield
\htmladdnormallink{E.Winstanley@sheffield.ac.uk}
{mailto:E.Winstanley@sheffield.ac.uk}
\end{center}

A half-day meeting entitled ``Brane-world gravity" was held at
Imperial College London on 13th November 2002, under the auspices of
the Gravitational Physics Group of the Institute of Physics.  The
meeting gave an overview of the state of the art in brane world
gravity, in a form accessible to non-experts and the many graduate
students present.

John March-Russell began with an overview of the physics of brane
worlds, including the latest experimental constraints, from colliders,
astrophysics and cosmology.  He concentrated on the phenomenology of
the simplest brane-world model, of Arkani-Hamed, Dimopoulos and Dvali,
in which the large extra dimensions are flat but compactified.  Ruth
Gregory then discussed gravity on the brane, including the
construction of brane-world cosmology geometries using patches of
Schwarzschild-anti-de Sitter space.  She also addressed the more
difficult (and as yet unsolved) problem of finding a metric which
describes a black hole on the brane.  Neil Turok's talk on the
cosmology of brane worlds prompted much lively debate.  He reviewed
recent work with Tolley and Steinhardt on the cyclic universes
resulting from colliding branes.

After a break for tea and discussion, Christian Trenkel described the
latest experimental work in Birmingham and elsewhere to test gravity
at short distances, in particular looking for the deviations from
Newtonian gravity predicted by brane-world models.  Marco Cavaglia
talked about the possibilities of black hole production at the next
generation of colliders, including the production cross-sections and
decay signatures.  Finally, Toby Wiseman described new pure gravity
solutions of low-energy string theory representing higher-dimensional
black holes, uniform homogeneous strings and non-uniform black
strings.

Many thanks to David Wands for organizing a very successful meeting.
\vfill
\eject

\section*{\centerline 
{Massive Black Hole Coalescence Focus Session at Penn State}}
\addcontentsline{toc}{subsubsection}{\it  
Massive black holes focus session, by Steinn Sigurdsson}
\begin{center}
Steinn Sigurdsson, Penn State
\htmladdnormallink{steinn@astro.psu.edu}
{mailto:steinn@astro.psu.edu}
\end{center}

The Center for Gravitational Wave Physics at Penn State ran a series of 
{\it focus sessions} in 2002 (see article by Sam Finn in Issue 19 of Matters
of Gravity, and \'Eanna Flanagan in Issue 20). The November 2002 focus session
was held the week of the 18th to the 21st, focused on the topic of the astrophysics
of the coalescence of supermassive black holes. The session was organised by
Ramesh Narayan, chair of the Center's advisory board,  Chris Mihos and myself.
Focus sessions are, by design, limited to a small number of invited attendees, 
and student applicants sponsored by their advisors. 
A web site with most of the presentations made at the meeting is at
\htmladdnormallink{http://cgwp.gravity.psu.edu/events/MBHMergers/}
{http://cgwp.gravity.psu.edu/events/MBHMergers/}.

The astrophysical processes that can remove energy and angular
momentum from a pair of supermassive black holes, bound to each other,
but embedded in a surrounding stellar population with significant
total mass, have been investigated to various levels of approximation
for many decades.  Despite a lot of research on the issues, the
fundamental problem of how supermassive black holes can transit from
the ``dynamical friction regime'' to the ``gravitational radiation
dominated regime'' remains intractable. The fundamental problem, as
noted by Begelman, Blandford and Rees (1980, see Quinlan 1996 for
historical review), is that once the BH binary becomes ``hard'',
further increases in binding energy is primarily achieved by ejecting
stars, which depletes the population of central stars available to
interact with the BH binary.  Relaxation processes can generally
refill the so-called ``loss-cone'' - the region in phase space where
particles can interact strongly with the binary, but only on very long
time scales. In simulations, the BH binary ``stalls'' at the ``last
parsec'', and additional physics is required if the binary is to
merge.

While BH binaries are observed (cf Komossa et al 2003, in fact the
announcement of the first confirmed detection of a bound binary
supermassive black hole was made, independently by NASA, during the
focus session), there is not an obvious, large population of tightly
bound binaries in the ``loss-cone'' regime, suggesting that there are
processes which enable rapid transit of the last parsec. Candidate
processes include rapid refilling of the loss-cone by dynamical
processes and hydrodynamical processes. There was also some discussion
of whether angular momentum extraction could be efficient, leading to
high eccentricity binaries (cf Aarseth 2002).

The session was split into a series of talks reviewing the physical
issues and discussing some recent results by some of the groups
working on the problem. Peter Bender, Doug Richstone and Tim de Zeeuw
reviewed the observational situation; Josh Barnes, Scott Tremaine and
myself reviewed some of the dynamical issues; Jeremy Goodman and
Ramesh Narayan reviewed the hydrodynamical issues and Martin Haehnelt
review the cosmological context and presented some new results on
estimated net merger rates in the context of hierarchical galaxy
formation. Qingjuan Yu, Pinaki Chatterjee, Priya Natarajan, Milos
Milosavljevic, Andres Escala, Kelly Holley-Bockelmann and Sverre
Aarseth presented some very interesting new results from recent
research.

The meeting finished with a very lively discussion led by Scott
Tremaine, and a review of the science and policy issues by Tom Prince.

Scott Tremaine summarised the work of the meeting, by preparing a list of issues that
are perceived to be decidable in by current research efforts.\\
{\it Decidable questions:}
\begin{itemize}
  \item{}Does the eccentricity of a hard binary BH increase or decrease?
  \item{}Is a given density of gas more effective than stars in merging binary BHs?
  \item{}Does triaxiality reduce the merger timescales for binary BH?
  \item{}Complete kinematic \& photometric data on all galaxies with measured BHs.
  \item{}Why do mergers of individual galaxies not give an NFW profile?
  \item{}What is the level of triaxiality in the centers of galaxies?
  \item{}What is the Brownian motion of a BH at the center of a cusp?
  \item{}Why are the orbits of stars near Sgr $A^*$ so eccentric?
  \item{}What are the limits on a massive binary BH in the Milky Way?
  \item{}What simulation method should we use for binary BH with gas?
  \item{}Can $M-\sigma$ relation be extended to higher and lower masses?
  \item{}Are intermediate mass BHs real?
\end{itemize}

\noindent And open issues of concern.\\

\begin{itemize}
  \item{}What is the evidence for binary BHs in AGNs?
  \item{}Is there any way to detect binary BHs in nearby inactive galaxies?
  \item{}Is there any way to detect BHs that are not at galactic centers?
  \item{}Will LISA tell us anything about quasars?
  \item{}Do we believe any N-body simulation?
  \item{}What is the gravitational radiation recoil in a BH merger?
\end{itemize}
 
See 
\htmladdnormallink{
http://cgwp.gravity.psu.edu/events/focus\_sessions.shtml}
{
http://cgwp.gravity.psu.edu/events/focus\_sessions.shtml}
 for past
and upcoming focus sessions.

We thank Ramesh Narayan and Chris Mihos for their efforts in
organizing the meeting.

{\bf References:}

Aarseth, S.J., ``Black Hole Binary Dynamics'', in ``Fred Hoyle's
Universe, eds. G. Burbidge, J.V. Narlikar and N.C. Wickramasinghe,
2003, \htmladdnormallink{astro-ph/0210116}
{http://arXiv.org/abs/astro-ph/0210116}.

Begelman, M.C., Blandford, R.D. \& Rees, M.J., Nature, 287, 307--309, 1980

Komossa, S., Burwitz, V., Hasinger, G., Predehl, P., Kaastra, J.S. \&
Ikebe, Y., ApJL, 582, L15--19, 2003

Quinlan, G.D., New Astronomy, 1, 35--56, 1996

\vfill
\eject
\section*{\centerline{ 
GWDAW 2002}}
\addcontentsline{toc}{subsubsection}{\it  
GWDAW 2002, by Peter Saulson}
\begin{center}
Peter Saulson, Syracuse University
\htmladdnormallink{saulson@physics.syr.edu}
{mailto:saulson@physics.syr.edu}
\end{center}

The seventh installment of the Gravitational Wave Data Analysis
Workshop was held in Keihanna Science City in Kyoto Prefecture, Japan,
on 17-19 December 2002. This meeting marked the rapid progress of the
various gravitational wave detection efforts around the globe. LIGO
Lab Director Barry Barish summarized the mood of most attendees in his
remarks to the workshop banquet, when he said that the meeting program
represented a mature field, with rapid progress on many instruments,
discussions of new analyses of unprecedentedly sensitive data, and
ongoing research on data analysis.

   The first day's talks were devoted to instrument progress
   reports. Interferometer reports all described substantial progress
   in commissioning and initial data taking runs. TAMA, GEO, and LIGO
   have all collected some science data in runs interspersed with
   commissioning aimed at achieving full design sensitivity. VIRGO has
   had a very successful commissioning exercise of its Central
   Interferometer, and will soon start commissioning its 3 kilometer
   arms. Attendees also learned of the vigorous prototyping program
   preparing for Japan's Large Cryogenic Gravitational Telescope
   (LCGT), slated for installation in the Kamioka mine. A 7-meter
   single-arm cryogenic test facility, CLIK, has been built at ICRR,
   while the LISM 20-meter room temperature interferometer is in
   operation at Kamioka. Construction of the 100-meter cryogenic CLIO
   interferometer has begun in the Kamioka mine; the tunnel has been
   prepared, and infrastructure work is now in progress. LCGT aims to
   take advantage of the low seismic noise of its underground site,
   and the low thermal noise of 20 K cryogenic test masses, to reach
   sensitivities sufficient to see neutron star binary inspirals at a
   distance of 200 Mpc. Advanced LIGO and others are also aiming at
   similar goals, although via different technological means.

   The growing excitement about the prospects of space-based detectors
   were the subject of the first afternoon's talks. LISA is at the
   center of the world's planning (it is a joint ESA-NASA project),
   but attendees also heard about a Japanese initiative called DECIGO,
   promoted by Seiji Kawamura, which is aimed at the 0.1 Hz band
   between LISA's most sensitive band and that of the ground-based
   interferometers. Among the bars, EXPLORER and NAUTILUS are
   operating well (more about them later), and ALLEGRO and AURIGA are
   about to come on line after major transducer upgrades.

   The second day's talks began the discussion of data analysis per
   se. The discussion was organized around signal character (burst,
   inspiral, sinusoidal, and stochastic), and moved fluently between
   bar and interferometer analyses. There were a number of talks from
   LIGO authors on the methods used to analyze data from the recent
   (late August to early September 2002) S1 run. Preliminary results
   are still embargoed to allow for revision during discussions
   internal to the LIGO Scientific Collaboration; most members of the
   LSC had only heard the first results of the analyses a week or two
   before the meeting, and an active review process is now under
   way. (The first big announcement of still-preliminary results is
   expected in mid-February at a meeting of the AAAS.)

The most-anticipated session of the meeting was the late-afternoon
section devoted to recent coincident analyses of data from
bars. Giovanni Prodi opened with a very clear discussion of the
methods used by the IGEC collaboration to set upper limits using
several years worth of data from the entire worldwide bar network. The
rest of the talks had as their subject the recent result from the Rome
group on 2001 data from EXPLORER and NAUTILUS (Astone et al.,
Class. Quant. Gravity 19, 5449-63 (2002) and \htmladdnormallink{gr-qc/0210053}
{http://arXiv.org/abs/gr-qc/0210053}.) In that
paper, "indications" were reported of the emission of gravitational
waves from sources scattered throughout the Galactic disk. Pia Astone
led off with a discussion of a Bayesian interpretation of the results,
(Eugenio Coccia having summarized the frequentist analysis of the
paper, during his status report the previous day.) Then came two
strong critiques of the statistical significance of the result,
presented in turn by Sam Finn and Warren Johnson. Finn demonstrated
that results as significant as those reported would be expected due to
chance alone 25\% of the time, hardly dramatic evidence of a
discovery. Astone, in remarks interspersed during Johnson's
presentation, emphasized that no discovery was claimed. Coccia closed
the session by defending the significance of the result, on the
grounds that detections at sidereal time 4 hours were special because
of the link to the Galaxy, but also reiterating that only further
observations could promote the claimed "indications" into a discovery
(or, of course, rule them out.)

Discussion spilled over into the evening and on to the workshop
banquet, aided by freely flowing sake and Asahi beer. Warren Johnson
and Eugenio Coccia mugged for the cameras, pretending to throw
roundhouse punches at one another. Then, Johnson and Pia Astone kissed
and made up, literally and repeatedly, until everyone's camera had
recorded the public reconciliation.

The third and final day was filled with talks on new data analysis
methods, discussions of sources, and accounts of detector
characterization techniques. While less easily summarized than the
previous two days' talks, these in some way were the most
future-oriented heart of the meeting, laying the foundations for the
new results to be expected in time for the next year's workshop.

On the sidelines of the meeting, another important discussion was
taking place. After the first day, representatives of TAMA and LIGO
met in the International Institute of Advanced Studies' beautifully
appointed seminar lounge to sign a Memorandum of Understanding for
joint analysis of data from the upcoming S2 data run. This brought to
fruition a series of negotiations spearheaded by Nobuyuki Kanda for
TAMA and Albert Lazzarini for LIGO. After the third and final day's
talks, a first working meeting of the two teams was held. Plans were
sketched for joint searches for burst signals and for chirps from
binary inspirals, intended to be completed six months after S2
concluded in mid-April 2003.

   Intensely-focused discussions of other ongoing or
   soon-to-be-initiated analyses were held during coffee breaks and in
   the evenings. Combined with the work formally presented in talks,
   these indicate a field truly mature, and ready for a steady stream
   of new scientific results over the next few years.

\vfill
\eject
\section*{\centerline{ 
Source simulation focus session at PennState}}
\addcontentsline{toc}{subsubsection}{\it  
Source simulation focus session, by Pablo Laguna}
\begin{center}
Pablo Laguna, PennState
\htmladdnormallink{pablo@astro.psu.edu}
{mailto:pablo@astro.psu.edu}
\end{center}

The Center for Gravitational Wave Physics at Penn State organized 
during October 28-30, 2002 a special "Focus Session" to address 
the topic of "Source Simulation and Gravitational Wave Data Analysis".
The goal of this focus session was the interplay between source simulations and 
gravitational wave data analysis: how the results of source simulations 
can be used to design data analysis that extracts more information, 
or information more efficiently, from gravitational wave detector data. 
The program dealt with the concrete, as opposed to the abstract: 
on developing analysis that make use of calculations relating 
to specific characteristics of specific sources.

The Program consisted of the following talks:

{\em ``Data Analysis in the Real World"} Sam Finn started his talk
reviewing the goals in source simulations and in data analysis. In
particular, he stressed the importance in carrying out source
simulations that identify the science reflected in the gravitational
waves.  For the data analyst, he stressed the need for developing
techniques that make science stand-out and provide astrophysical
interpretation of observations. He also review the different classes
of sources: stochastic, periodic and bursts.

{\em ``Linguistics of LISA Sources"} Scott Hughes presented an
overview of key LISA sources. In particular, he discussed questions
such as: What do we hope to learn from LISA sources? What is the
character of the signals these sources generate?  He also addressed
the issue of how to design a strategy to measure LISA sources.  He
pointed out that there is a big difference between detection and
measurement.  He also discussed whether we can combine GW information
with other channels to maximize the astrophysical payoff.

{\em ``Structure, Stability and Dynamical Behavior of Compact
Astrophysical Objects"} Joel Tohline presented a review of a meeting
that took place the weekend before the workshop. This meeting was
focused on two types of mechanisms in instabilities of compact
objects: Hydrodynamical instabilities such as bar-mode instabilities,
and GR driven instabilities (e.g. r-modes). Some of the issues
discussed in this meeting were: Mode identification, damping
mechanisms, expected maximum amplitude and duration, effects from GR
on mode character and likelihood of producing detectable GW signals.

{\em ``Gravitational Waves from the Tidal Disruption of Neutron Stars
in Binaries"} Michele Vallisneri emphasized the importance of
investigating the correspondence between the EOS and mass-radius
function of NS. He discussed the possibility of using information from
NS tidal disruption in NS-BH binaries as a probe of the NS EOS.

{\em ``Gamma Ray Bursts and Gravitational Waves"}
After a short review of GRBs, Shiho Kobayashi discussed 
how the detection of counterparts of GRBs in GWs could
revolutionize the field. He pointed out that one 
can use GRBs and afterglow observations 
about the time and location of the event to perform a
cross-correlation and obtain information of the association
between GRBs and gravitational wave bursts.

{\em ``Predicting the Gravitational Wave Signatures of Core Collapse
Supernovae: The Road Ahead"} Tony Mezzacappa presented the road
required in order to solve the core collapse supernovae problem,
including the gravitational radiation produced by these events. He
pointed out that waveforms will not be available any time soon. This
is an extremely difficult problem requiring a 3D-GR-Radiation-MHD code
plus state of the art nuclear and weak interaction physics.  His talk
provided an overview of the current state of these simulations.  In
particular, he showed simulations of accretion shock instabilities and
neutrino driven convection.

{\em ``Gravitational Waves from Supernova Core Collapse: What Could
the Signal Tell Us"} Harald Dimmelmeier presented results from
simulations of supernova core collapse. He reviewed the physical
complexity and numerical difficulties involved in relativistic
simulations of rotational core collapse to a neutron star. He pointed
out that because of these complications it is necessary to introduce
several approximation.  However, these approximations do not prevent
us from extracting new physics encoded in the waveforms. Their
simulations show that the remnants are more compact with higher
densities when compared with Newtonian results. In addition,
relativistic effects seem to increase the rotation rate, and in many
instances these effects could trigger tri-axial instabilities.

{\em ``Low Frequency Gravitational Waves from the Galactic Halo"}
Shane Larson gave a talk reviewing first the MACHO search. 
He then discussed the potential for LISA observations of 
gravitational radiation from white dwarfs and black hole
MACHO binaries.

{\em ``Binary Black Hole Coalescence in Galaxy Mergers"}
Steinn Sigurdsson stressed that although BBH coalescences
in galaxy merger could potentially have large S/N, the
rate of these events is likely to be low. He also addressed E\&M and Spin
signatures as well as the possibility of observing stars bound to the BHs. 

Other talks in the meeting included: ``Gravitational Wave Observations
of Galactic Populations of Compact Binaries" by Matthew Benacquista,
``An Overview of 3D Black Hole Simulations" by Pablo Laguna,
``Bothrodesy: The Promise and Challenges of Extreme Mass Inspirals''
by Teviet Creighton, ``Probing the Equation of State of Neutron Stars
with LIGO" by Fred Rasio

Links to presentations as PDF files can be found at\hfill

\htmladdnormallink{
http://cgwp.gravity.psu.edu/events/SrcSimDA/}
{
http://cgwp.gravity.psu.edu/events/SrcSimDA/}

\vfill
\eject
\section*{\centerline{ 
Raman Research Institute Workshop}\\
\centerline{on Loop Quantum
Gravity}}
\addcontentsline{toc}{subsubsection}{\it  
RRI workshop on loop quantum gravity, by Fernando Barbero}
\begin{center}
J. Fernando Barbero G., IMAFF-CSIC, Madrid
\htmladdnormallink{jfbarbero@imaff.cfmac.csic.es}
{jfbarbero@imaff.cfmac.csic.es}
\end{center}

During two weeks in November and December 2002, approximately 20
researchers in quantum gravity gathered at the Raman Research
Institute in Bangalore for an intensive meeting on Loop Quantum
Gravity and related issues. Talks were divided into morning
sessions, shared by Abhay Ashtekar and Martin Bojowald, and
afternoon sessions for the rest of the participants.

Abhay Ashtekar gave quite a detailed overview of three main
topics: Quantum geometry, black hole entropy, and semi-classical
issues. After an introductory talk to set the stage of his series
of lectures and a discussion on the quantum mechanics on SU(2), he
gave two talks about connections on graphs and field theories of
connections, both from the classical and the quantum perspective.
They were followed by a lecture devoted to quantum Riemannian
geometry and the definition of area and volume operators. After
his introduction on quantum geometry he talked about the quantum
geometry of isolated horizons and black hole entropy; he
discussed, in particular, new results on the computations of
entropy for non-minimally coupled scalar fields. Finally his three
last talks were devoted to the discussion of semi-classical
issues. In the first he used the quantum mechanics of a particle
to study how the polymer-like quantum theory  of a particle
reduces to the usual Schr\"{o}dinger Quantum Mechanics in the low
energy regime. He then considered Maxwell theory in the next
lecture, complementing a previous discussion of Madhavan
Varadarajan's first talk and finished with an enlightening lecture
on the relationship between the Fock, r-Fock, and polymer
representations of Maxwell theory. On Dec, 4th he gave a public
Academy Lecture on loop quantum gravity in front of a full
auditorium. The ten one-and-a-half hour talks succeeded in
providing an introduction to the main topics for non-experts and
yet gave an insightful review of the present state of loop gravity
and its applications. The talks sparked a lot of interaction among
the participants in coffee breaks and informal discussion sessions
and gave a clear idea of the status of the program and its future
development. In spite of a winter cold Abhay displayed a lot of
enthusiasm and energy that were certainly inspiring for the rest
of the participants.

Martin Bojowald's talks we devoted to the mathematical issues
related to symmetry reductions of theories of connections and
their application to the study of loop quantum cosmology. He
devoted his first three talks to the discussion of the mathematics
of symmetry reductions for theories of connections and the
definition of symmetric states in quantum geometry. After that he
gave a thorough introduction to the kinematics of cosmological
models, matter Hamiltonians, quantization ambiguities, and the
study of homogenous and isotropic models. He discussed several
cosmological issues from the loop quantum gravity perspective; in
particular he gave a detailed overview of the meaning of the
initial singularity and initial conditions, evolution through
classical singularities, large volume behavior and corrections to
the classical approach to singularities. He also discussed the
possibility of explaining inflation within his framework. It was
really impressive to see how far the developments on loop quantum
cosmology have reached and the impulse that Martin's work is
giving to the whole loop quantum gravity program.

Afternoon sessions covered a much broader set of topics on
classical and quantum gravity, cosmology, and quantum field
theory. Talks were shorter and were given by a number of speakers.
Sukanya Sinha reviewed the approach  to the semiclassical Einstein
equations pioneered by Bei Lok Hu and to which she has made recent
contributions.  P. Majumdar discussed some aspects related to the
computation of quantum corrections to black hole entropy, in
particular he discussed computations for BTZ and AdS-Schwarzschild
black holes. L. Sriramkumar gave two introductory talks discussing
some issues related to inflation and quantum gravity. T. R.
Govindarajan gave us a talk about recent developments in de Sitter
gravity, specifically on the thermodynamics of the cosmological
horizon of the de Sitter space-time. J. Samuel discussed a novel
proposal to test general relativity by using radio interferometers
with intercontinental baselines and measuring the curvature of the
wavefront emitted by a distant source. G. Date gave a talk on a
recent model of his to describe discrete time evolution in a
simple quantum mechanical system that mimics some of the features
of the symmetric cosmological models discussed by Bojowald.
Madhavan Varadarajan gave several talks about his last work on
quantum linearized gravity and the r-Fock representation, in
particular the use of U(1) flux nets in the context of the simple
model provided by electromagnetism. Finally, the author of these
lines discussed his work on diff-invariant and non diff-invariant
free actions and their use in perturbative quantum gravity.

The meeting was very successful in all the possible aspects. The
talks were interesting and illuminating with a lot of discussion
during and after the seminars. The atmosphere among the
participants was excellent and the scientific exchange really
fruitful. I would like to emphasize the perfect organization of
the meeting by Madhavan and the very warm hospitality of the whole
theory group at RRI. I believe that all the participants are
looking forward to attending future meetings at the Raman
Institute.

\vfill
\eject
\section*{\centerline {Lazarus/Kudu Meeting at PennState}}
\addtocontents{toc}{\protect\smallskip}
\addcontentsline{toc}{subsubsection}{\it  
Lazarus/Kudu Meeting, by Warren G. Anderson}
\begin{center}
    Warren G. Anderson, The University of Texas at Brownsville\\
\htmladdnormallink{warren@phys.utb.edu}
{mailto:warren@phys.utb.edu}
\end{center}

On Sept. 7-9, 2002, the Center for Gravitational Wave Physics (CGWP)
at Penn State hosted a small informal meeting to discuss the results
of the Lazarus project and their relationship to the detection of
gravitational waves from merging black holes. This meeting, organized
by Manuela Campanelli, Sam Finn (CGWP director) and Pablo Laguna, was
designed to maximize the rate of information exchange by limiting
participation to just a few people, most of whom were already active
in the Lazarus and Kudu projects. In attendance were Abhay Ashtekar,
John Baker, Bernd Bruegmann, Jordan Camp, Manuela Campanelli, Joan
Centrella, Jolien Creighton, Sam Finn, Pablo Laguna, Ben Owen, Deirdre
Shoemaker and I.

For those not familiar with the Lazarus project, it is an attempt to model
black hole mergers using the limited available numerical evolution only when
absolutely necessary. The goal of the project, led by John Baker, Manuela
Campanelli and Carlos Lousto, is to seamlessly sew a complete 3+1D, nonlinear
numerical evolution in between a suitable early time approximation of the
binary system, such as the PN approximation, and a suitable late time
approximation, such as the close limit approximation, and thus produce merger
waveforms which are suitable for data analysis. At present, the waveforms
produced by Lazarus are not of sufficient accuracy to use for the standard
data analysis algorithm of matched filtering. This has led to a second
project, the Kudu project, whose goal is to define a framework for going from
imperfect waveforms to data analysis algorithms, and to implement the
framework on the Lazarus waveforms.

As with many ``working meetings'' that are being hosted by the CGWP, the
agenda featured brief talks followed by long discussion periods, a format
which I like very much. The first day of the meeting focused on the machinery
of the Lazarus approach. Manuela Campanelli started with an overview of the
Lazarus and Kudu projects. Almost immediately, there was a flurry of questions
and discussion which set the tone for the day - a detailed, highly interactive
dissection of the Lazarus project. Carlos Lousto's presentation of Lazarus
methodology and John Baker's talk on Lazarus results were direction markers
which primarily served to guide the general course of the conversation in
which we were already engaged. It would be futile to try to detail the flow of
this freeform discussion. Let me instead summarize with what I think was the
main conclusion of the day - the Lazarus project has developed a framework
which must overcome many technical hurdles, and at each of these hurdles,
there is room for error to enter into the calculation. At present, much of the
knowledge about these errors has been gleaned by the Lazarus researchers
project seeing how things go awry when a mistake is made or they adjust a
parameter. However, in order to use these waveforms, data analysts will need
quantitative error estimates. Providing these estimates will be a long and
arduous process, but the Lazarus project members who were present seemed to
agree that it was possible and worthwhile.

The second day of the meeting focused on the Kudu project. This day was much
more speculative - the Kudu project had barely started when the meeting was
held. As a result, the nature of the meeting changed from critical review to
exploring new horizons. I gave the first talk of the day, in which I described
how one could devise an optimal (in a certain sense) search algorithm for
signals about which only incomplete information was available. Much
interesting discussion followed on how one could use this to help refine
source modeling efforts by concentrating the effort on those aspects of the
model whose refinement would lead to the greatest increase in detection rate.
Jolien Creighton then outlined the methods that are currently being used
within LIGO to search for unmodeled signals, and how these methods might be
improved to look for the Lazarus waveforms.

On the final morning, we had a discussion of where we stood and future
directions. However, as with all such discussions, it is in the doing that
progress will be made, rather than in the discussion. And this leads me to
what I believe was the most fruitful aspect of this meeting. Over coffee,
dinners, and late into the evenings, a subgroup of us discussed specific ideas
that we wanted to explore, and agreed to have weekly teleconferences to
discuss the results we have obtained. This group continues to meet (biweekly
now) and to discuss results obtained and brainstorm new directions. It is
unlikely that this nucleus could have self-assembled without the environment
of a working meeting to stimulate it, and it is exactly by providing such
opportunities that I feel that the CGWP will have its greatest impact on our
community.

\end{document}